# A STUDY OF CROSS-CORRELATION AND BREAKFINDING ALGORITHMS APPLIED TO THE MEASUREMENTS OF REDSHIFTS IN VERY LOW RESOLUTION SPECTRA

R. Cabanac and E. F. Borra

Centre d'Optique, Photonique et Laser, Observatoire du Mont Mégantic
Département de Physique, Université Laval,
Québec, QC
CANADA G1K 7P4
Electronic mail: cabanac@phy.ulaval.ca, borra@phy.ulaval.ca

POSTAL ADDRESS: Département de Physique, Université Laval, Québec, QC, Canada G1K 7P4

**ABSTRACT**

In this article we compare the cross-correlation and breakfinder techniques applied to the measurements of redshifts from low-resolution spectra. We assume spectra obtained from multinarrowband imagery, a technique for multi-object spectrophotometry. Comparing the cross-correlation with the breakfinder, we find that neither is intrinsically superior to the other. They have comparable precision for early type galaxies but the cross-correlation is clearly superior for later type galaxies. On the other hand, unlike the breakfinder, the cross-correlation is sensitive to instrumental errors as well as the shape of the templates used. We carried out simulations with two types of global spectral errors: a slope error and sinusoidal errors. They can introduce serious systematic errors in redshifts measured with the cross-correlation. The breakfinder is a spectrally local test and does not suffer from this type of error. We therefore conclude that the two techniques are complementary and that they should both be used to help flag objects for which they give abnormally discrepant redshifts.

## 1. Introduction

Cosmological objects tend to be faint and cosmological studies tend to be statistical in nature: There thus arises the need for observations of large numbers of very faint objects. A practical way to do this kind of work is to use either slitless spectroscopy or filter photometry that yield very low-resolution spectroscopic information. Redshifts give crucial cosmological information and one should wonder how accurately they can be determined from low resolution spectrophotometry. Attempts to obtain approximate redshifts from low-resolution spectrophotometry are not new. For example, Baum (1962), Coleman, Wu & Weedman (1980), Koo (1985) and Loh & Spillar (1986) have used wide-band filters to get redshifts. One of the problems with a filter technique using wide bands is that it is not very accurate.

Correlation techniques with templates have been tried by Oke (1971). They can measure redshifts and also classify the galaxy by its spectral type (Hickson, Gibson & Callaghan 1994). The 4,000 Å break is the strongest feature detectable in low-resolution spectra of early type galaxies and contains most of the redshift information. An interactive technique to measure redshifts with the 4,000 Å discontinuity from low-dispersion spectra has also been developed by Beard et al. (1986) and Cooke et al. (1986). An automated technique to measure redshifts from the 4,000 Å break in slitless spectra has been discussed by Borra & Brousseau (1988) and used by Beauchemin & Borra (1994).

In practice, spectrophotometry with interference filters is superior to slitless spectroscopy for the observation of very faint objects (Hickson, Gibson & Callaghan 1994). Hickson, Gibson & Callaghan (1994) have carried out simulations to determine redshifts with the cross-correlation technique applied to noise-degraded spectral energy distributions of galaxies.

In this article ( see Cabanac 1992 for more detailed results) we apply the cross-correlation and breakfinder techniques to determine redshifts from low-resolution spectra and compare the performance of the two techniques. We assume that spectra are obtained from multinarrow band imagery with interference filters 200 Å to 300 Å wide. This kind of data is of current interest since it is presently being obtained from a survey that uses a 2.7-m liquid mirror telescope (Hickson et al. 1994). A deeper survey will later be carried out with a 5-m liquid mirror telescope presently under construction by a collaboration between the University of British Columbia, the Institut d'Astrophysique and Laval .

## 2. REDSHIFTS FROM NOISE-DEGRADED LOW-RESOLUTION SPECTRA

We start from a library of 5 galaxy spectra (Pence 1976) that we redshift from $z = 0.005$ to $z = 0.6$ in steps of $\Delta z = 0.005$. Cabanac (1992) obtained similar results by using other spectral catalogs. The resulting 600 spectra are then sampled with the set of 40 interference filters of the UBC-Laval survey (Hickson et al. 1994). They have central wavelengths spaced uniformly with logarithm in the range 4,000 to 10,000 Å. The spectra are then noise-degraded with the help of a random number generator for signal to noise ratios (S/N) per filter of 100, 50, 20, 10, 5, 3. To have sufficient statistics, we generated 100 noisy spectra per redshift interval per galaxy type for each of the 6 S/N for a total of 360,000 spectra. We then analyzed the spectra with the breakfinding technique (Borra & Brousseau 1988) and the cross-correlation technique (Hickson, Gibson & Callaghan 1994).

a). Redshifts from the 4,000 Å break

The strongest features detectable in low-resolution spectra of early type galaxies are the 4,000 Å break and another feature caused by the red edge of the G band having a rest wavelength of about 4,440 Å. They contain most of the redshift information. The breaks are quite strong in the spectra of elliptical galaxies as well as in Sa and Sb spirals but are weaker in later type spirals (Coleman, Wu & Weedman 1980). Previous studies have shown that neither the contrast nor the rest wavelength of the 4000 Å break changes significantly when z<0.5 (Dressler 1987, Hamilton 1985).] A break-finding algorithm has been developed by Borra & Brousseau (1988) that finds directly, and without human intervention, the position of spectral discontinuities in spectra. Briefly, the algorithm creates 2 spectral windows having equal width *a*. They are separated by an "opaque" window containing *b* pixels ( an odd number). The software then moves the windows across the spectrum and computes the flux differences. The function *Cj*, computed for every pixel *j*, is defined by

$$C_j = \sum_{i=j+(b+1)/2}^{j+a+(b-1)/2} I_i - \sum_{i=j-a-(b-1)/2}^{j-(b+1)/2} I_i \qquad (1)$$

*Cj* is then smoothed with a gaussian to facilitate the detection of the true maximum of *Cj* in faint objects for which a noisy energy distribution gives too many peaks of *Cj*. The width of the gaussian is found after experimentation. *Cj* is proportional to the first derivative of the spectrum and the nominal position of the spectral break is thus given by the peak of *Cj* which is the point at which the second derivative of the spectrum is zero. The nominal position of a spectral break is thus given by the point at which the slope changes sign. The maximum of *Cj* is first determined to the nearest integer pixel by finding the filter that has the highest value of *Cj*. Subsequently, the fractional filter position at which *Cj* is maximum is determined by least squares fitting a parabola centered on the integer pixel peak of *Cj*.

The standard deviation of *Cj* is readily obtained from Equation 1 as

$$\sigma_j^2 = 2 \sum \sigma_i^2 \quad , \qquad (2)$$

where $\sigma_i$ is the standard deviation of the intensity $I_i$ at pixel i and the limits of the sums are the same as in Equation 1. One must experiment with several values of *a* and *b* to find the optimum values for a given resolution and dispersion. For our filter system, we found that the optimum value of *a* was two "pixels" ( actually 2 filters in our case) wide, corresponding to 400 to 600 Å and b was one "pixel" wide (200 to 300 Å).

The algorithm has been shown to work well with actual photographic slitless spectra that had resolutions of about 100 Å and low S/N (<10/pixel). Borra & Brousseau (1988) applied the break finding algorithm to noisy slitless spectra of galaxies in a cluster of galaxies , finding standard deviations $\sigma \sim 3{,}000$ Km/sec. Beauchemin & Borra (1994) successfully detected redshift peaks corresponding to large scale structures independently found by others.

To estimate the performance of the break-finding algorithm for our case, we applied it to the 360,000 noise degraded spectra. For every galaxy type and S/N, we have counted the number of objects detected for a given redshift standard deviation value. The results are shown in figures 1 for E and S0 galaxies, 2 for Sab galaxies and 3 for later type spirals. They give the frequencies of detection within a given redshift error and can readily be used to estimate the redshift precision as function of Hubble type and signal to noise ratio. For example, figure 1 shows that, at S/N=10, 70 % of redshifts of Ellipticals and S0s are obtained within 1,200 Km/sec of the true values so that 1,200 Km/sec corresponds to about a standard deviation. Figure 2 shows that the standard deviation increases to 3,000 Km/sec for Sabs and figure 3 that it increases to 6,000 Km/sec for later type spirals. The precision increases with S/N.

One of the main sources of uncertainties of the breakfinding algorithm comes from difficulties in identifying the 4,000 Å break from other spectral features. This problem is not as serious for early type galaxies that have conspicuous 4,000 Å breaks but it worsens for late type spirals that have much weaker discontinuities. The difficulty becomes quite serious for all Hubble types observed with S/N <10. The redshift precision would improve if one could find another technique that would give a redshift sufficiently accurate to preclude the confusion with other spectral features.

b) redshifts from cross-correlations

The breakfinder gives redshifts but does not determine the Hubble type. On the other hand, the cross-correlation technique (Tonry & Davis 1979) can determine redshifts and the spectral type. Let $S(\lambda)$ be the spectrum of an object and $P(\lambda)$ the correlation template. Then the cross-correlation vector is given by

$$C_j = \frac{\sum_{i=1}^{No\ filters} S_i P_{ij}}{n \sigma_s \sigma_p} \quad (3)$$

$$\sigma_s = \left[ \frac{\sum_{i=1}^{No\ filters} S_i^2}{n} \right]^{1/2} \quad (4)$$

$$\sigma_p = \left[ \frac{\sum_{i=1}^{No\ filters} P_i^2}{n} \right]^{1/2} \quad (5)$$

The correlation is performed by successively fitting all of the galactic templates $P(z,\lambda)$ redshifted from 0 to 0.6. The goodness of fit of a template is then obtained from equations 3 to 5. The best fitting template gives the Hubble type and the redshift. The templates were generated the same way as for the breakfinder technique. The same numbers of noise-degraded templates were used.

We find that the success rates of the identifications of the Hubble types differ little among our E-S0, Sab, Sbc, Scd and Sm-Im classes and typically are 100% for S/N≥20, >95% for S/N= 10 and >70% for S/N= 5. This is in agreement with the findings of Hickson, Gibson & Callaghan (1994).

Figures 4 to 8 are the equivalent of figures 1 to 3 but for the cross-correlation technique. Comparing the 2 sets of figures, it appears that, although the breakfinder has better performance for S/N <10, the 2 techniques are essentially equivalent for E-S0 and Sab; however the cross-correlation technique is somewhat better for Sbc and gives results for later Hubble types for which the breakfinder does not work at all. However, the breakfinder has the advantage that it is a local technique. As such, it gives results even with a reduced number of filters, at least in a restricted redshift interval. We have carried out simulations with the cross-correlation technique for reduced numbers of filters, finding that the errors increase noticeably below 20 filters. Furthermore, the breakfinder has the advantage of not being sensitive to the effect of global features such as those introduced by reddening (either from the galaxy itself or our galaxy) or instrumental effects.

The cross-correlation is sensitive to global spectral departures from the templates. Such departure can arise from natural galaxy to galaxy differences, reddening or instrumental (e.g. flat fielding) effects. We have investigated the effects of global features on redshifts obtained from cross-correlations. We have imposed two types of global features: an error in the slope of the continuum and sinusoidal errors superposed on the continuum.

The importance of reddening can be roughly estimated by considering that the global shapes among the spectra of the various Hubble types only differ from one another at the 20 % level (e.g. Sbc from Scd in the range 5000 Å to 1 micron). Reddening in the Galaxy roughly goes as the inverse of the wavelength in optical spectrum ($A_v$ ~ 3) and also varies with the optical thickness of the disk. This reddening changes the slopes of galaxy spectra particularly in the 0.5um-1um range by

$$\Delta(\text{slope})/\text{Slope} = \{I(\lambda_1) - I(\lambda_o) - [I(\lambda_1) - dI(\lambda_1) - I(l\lambda_o) + 2dI(\lambda_o)]\}/ I(\lambda_1)-I(l\lambda_o), \quad (6)$$

and therefore

$$\Delta(\text{slope})/\text{Slope} = d \{ 1 - I(\lambda_o) / [I(\lambda_1)-I(\lambda_o)] \}, \quad (7)$$

where $I(\lambda_1)$, $I(\lambda_o)$ are the fluxes at $\lambda_1$ and $\lambda_o$ ($\lambda_1 = 2\lambda_o$) and the absorption is twice as large at $\lambda_1$ as at $\lambda_o$, d is a coefficient of absorption (%) by the interstellar medium (ISM). This simple analysis shows that the ISM extinction can induce significant changes in the continuum since a 10% absorption at $\lambda_1$ (20% at $\lambda_o$) could easily introduce a 20% change in the slope of the continuum. Moreover, we can assume that similar effects occur in galaxies showing a similar galactic type. The accumulation of exogenous and endogenous effects are both unpredictable and effective in changing the slope of our spectra.

Let us first consider the effect of errors on the slope of the continua. Figure 9 shows the effect of a 10% slope error on spectra having S/N =100. It plots the residuals for all types for 0.0<z<0.6. Figure 9a shows the original data, while 9b gives the residuals for the 10% slope error and 30 filters and 9c for 10% error and 20 filters. We can see that there are systematic effects of the order of 1,000 to 5,000 Km/sec for some types and some redshift intervals. The performance is somewhat worse for 30 filters than 20 filters.

We have also made simulations with a sinusoidal error added to the continua. Sinusoidal errors approximate higher frequency instrumental errors (e.g. from poor rectification or local sky subtraction). This was done for several periods and amplitudes. A 1% amplitude gives small errors that are barely noticeable (<1,000 Km/sec ). However a 10% error has important consequences. Figure 10a,b,c is the equivalent of Fig. 9 but for a sinusoidal error having a 10% amplitude and respectively, a period of 6,000 Å, 3,000 Å and 1,500Å. The correlation was carried out with 30 filters. The sinusoidal errors with long periods affect colors, while those with shorter periods affect more local features. The discontinuities are caused by mismatched galactic types. The discontinuities disappear if we only use 20 filters. Figures 11 to 15 show the effects of a sinusoidal error having 10% amplitude and a period of 6,000 Å on the detected Hubble type as functions of S/N. Given a Hubble type, each figure gives the Hubble type identified by the cross-correlation.The errors are non-negligible, especially for Sabs, even at high S/N.

It would thus appear that sinusoid-like errors in the continuum introduce non-negligible errors. Therefore one must take great care to avoid these types of errors. They should be kept at the 1% level or below. Ideally one should suppress the continuum and only keep local features. Tonry & Davis (1979) used the Fourier transform to eliminate this kind of errors; unfortunately Fourier transforms are computing time intensive and ill-suited to the analysis of large surveys that generate massive quantities of data. It may however be possible to eliminate the continuum, with reasonable computing times by heavily smoothing it and then subtracting it from the spectrum. To eliminate the continuum, we have performed the correlation with the derivative of the spectrum, finding that 10% amplitude sinusoidal errors have no noticeable effects. Unfortunately the redshift errors increase faster with decreasing S/N than is the case for the redshifts obtained from cross-correlation of the continuum ,the expected consequence of taking a derivative of the spectrum. We are thus faced with the dilemma of having to choose between freedom of systematic effects and better performance at low S/N.

## 3. CONCLUSION

Comparing the crosscorrelation and the breakfinding techniques, we find that neither is intrinsically superior to the other. They have comparable precision for early type galaxies but the cross-correlation is clearly superior for later type galaxies. This is however not a serious handicap since the breakfinder would perform as well as the cross-correlation for about half of all galaxies. On the other hand, the breakfinder has the important advantage of spectral locality and, unlike the cross-correlation technique, is not sensitive to mismatching of the template. We carried out simulations with two types of global spectral errors: a slope error and sinusoidal errors. We find that they can introduce serious systematic errors in redshifts measured with the cross-correlation. The slope error is potentially more serious than sinusoidal errors since the latter can be kept sufficiently low with proper data processing. On the other hand, reddening (either in our galaxy or internal to the galaxies studied) is more worrisome since it can easily introduce important slope errors. We therefore conclude that the two techniques are complementary, should both be used and help to flag objects for which they give abnormally discrepant redshifts.

## ACKNOWLEDGMENTS

R. Cabanac was supported by a Natural Sciences and Engineering Research Council of Canada graduate fellowship. This research was supported by a Natural Sciences and Engineering Research Council of Canada grant to E.F.B.

FIGURE CAPTIONS

Fig. 1:

Percentage of objects detected with the breakfinding algorithm for a given redshift standard deviation value for E and S0 galaxies. The figure gives the frequencies of detection within a given redshift error.

Fig. 2:

Percentage of objects detected with the breakfinding algorithm for a given redshift standard deviation value for Sab galaxies. The figure gives the frequencies of detection within a given redshift error.

Fig. 3:

Percentage of objects detected with the breakfinding algorithm for a given redshift standard deviation value for later type spirals. The figure gives the frequencies of detection within a given redshift error.

Fig. 4:

Percentage of objects detected with the cross-correlation for a given redshift standard deviation value for E and S0 galaxies. The figure gives the frequencies of detection within a given redshift error.

Fig. 5:

Percentage of objects detected with the cross-correlation for a given redshift standard deviation value for Sab galaxies. The figure gives the frequencies of detection within a given redshift error.

Fig. 6:

Percentage of objects detected with the cross-correlation for a given redshift standard deviation value for Sbc galaxies. The figure gives give the frequencies of detection within a given redshift error.

Fig. 7:

Percentage of objects detected with the cross-correlation for a given redshift standard deviation value for Scd galaxies. The figure gives the frequencies of detection within a given redshift error.

Fig. 8:

Percentage of objects detected with the cross-correlation for a given redshift standard deviation value for Sm-Im galaxies. The figure gives give the frequencies of detection within a given redshift error.

Fig. 9:

It shows the effect of a 10% slope error on redshifts obtained with the cross-correlation for spectra having S/N =100 per pixel. It plots the residuals for all types for 0.0<z<0.6. Figure 9a shows the original data, while 9b gives the residuals for the 10% slope error and 30 filters and 9c for 10% error and 20 filters.

Fig.10:

It shows the effect of a sinusoidal error having a 10% amplitude and periods of 6,000 Å (10a), 3,000 Å (10b) and 1,500Å (10c) on redshifts obtained with the cross-correlation for spectra having S/N =100/pixel. The correlation was carried out with 30 filters. The sinusoidal errors with long periods affect colors, while those with shorter periods affect more local features.

Figure 11:

It shows the effects of a sinusoidal error having 10% amplitude and a period of 6,000 Å on the detected Hubble type as functions of S/N. Given an E-So, the figure gives the Hubble types identified by the cross-correlation.

Figure 12:

It shows the effects of a sinusoidal error having 10% amplitude and a period of 6,000 Å on the detected Hubble type as functions of S/N. Given a Sab, the figure gives the Hubble types identified by the cross-correlation.

Figure 13:

It shows the effects of a sinusoidal error having 10% amplitude and a period of 6,000 Å on the detected Hubble type as functions of S/N. Given a Sbc, the figure gives the Hubble types identified by the cross-correlation.

Figure 14:

It shows the effects of a sinusoidal error having 10% amplitude and a period of 6,000 Å on the detected Hubble type as functions of S/N. Given a Scd, the figure gives the Hubble types identified by the cross-correlation.

Figure 15:

It shows the effects of a sinusoidal error having 10% amplitude and a period of 6,000 Å on the detected Hubble type as functions of S/N. Given a Sm-Im , the figure gives the Hubble types identified by the cross-correlation.

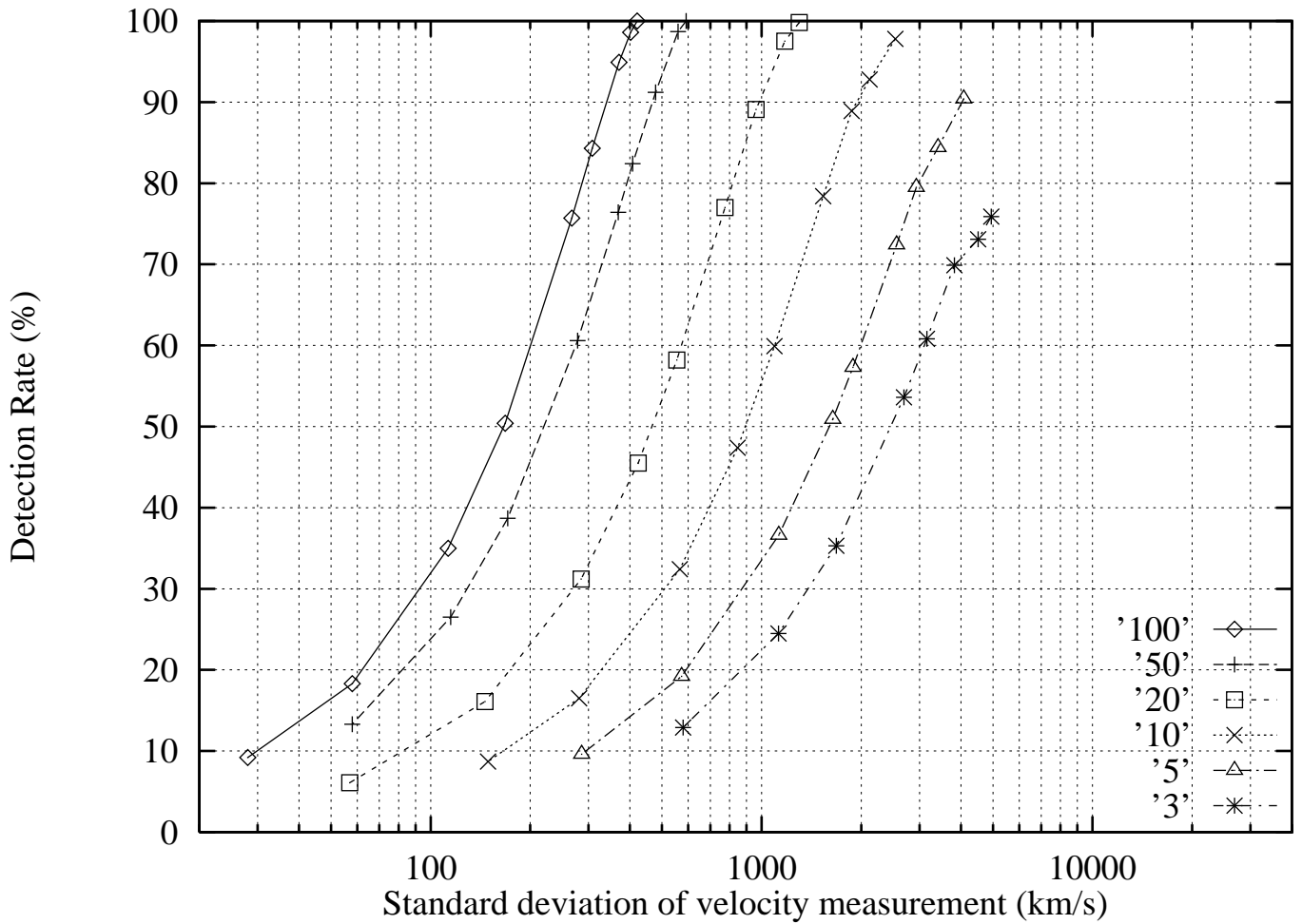

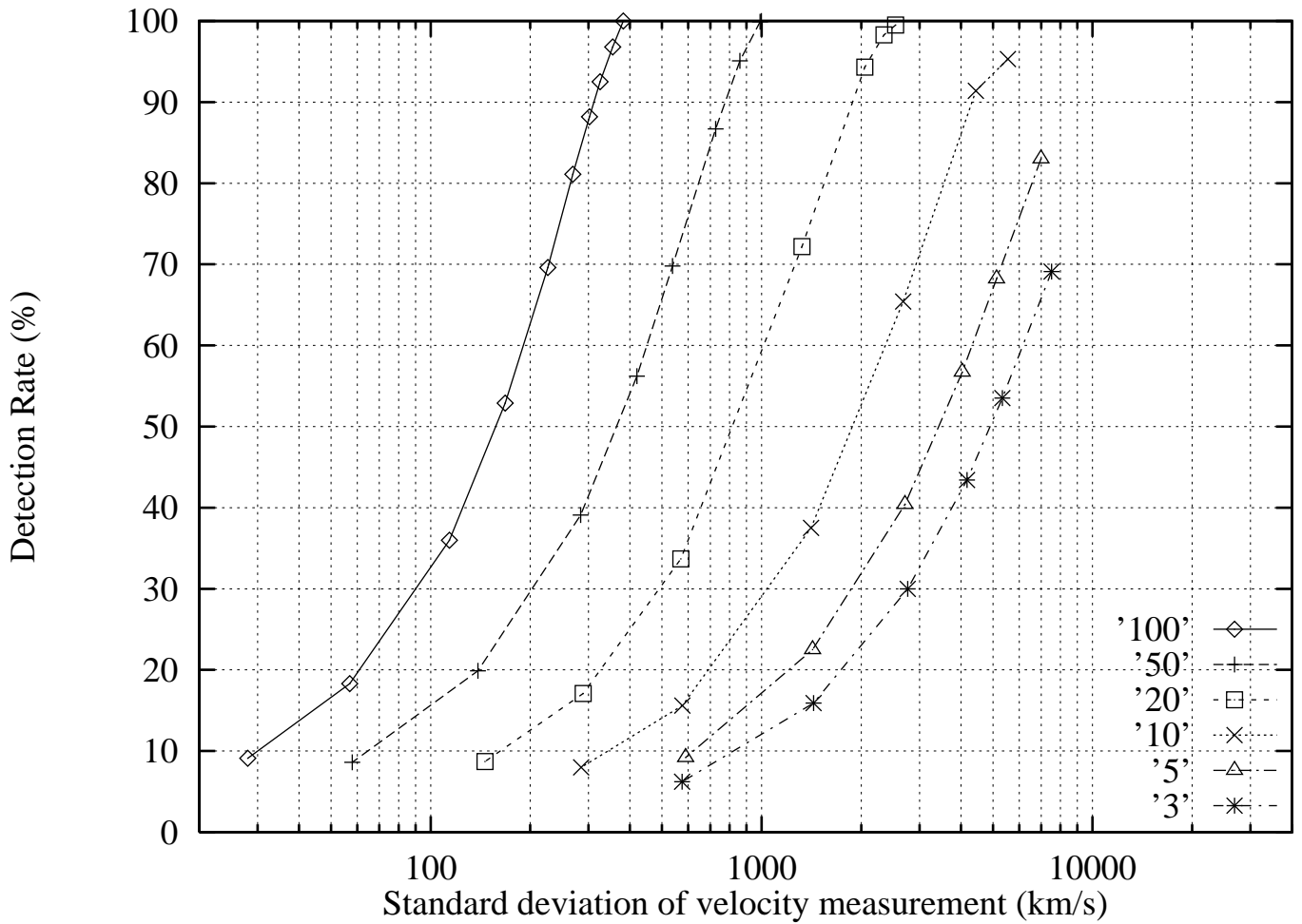

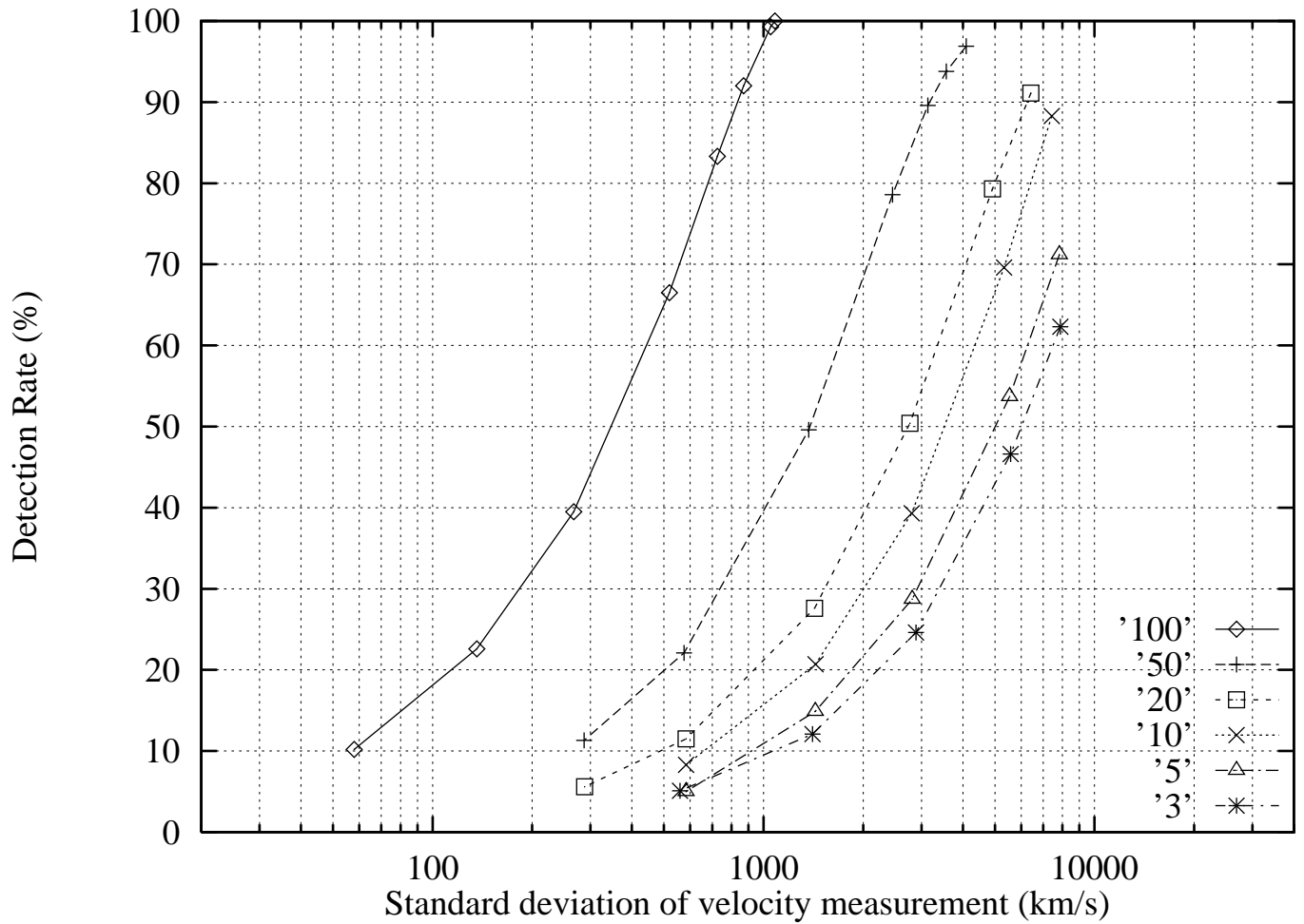

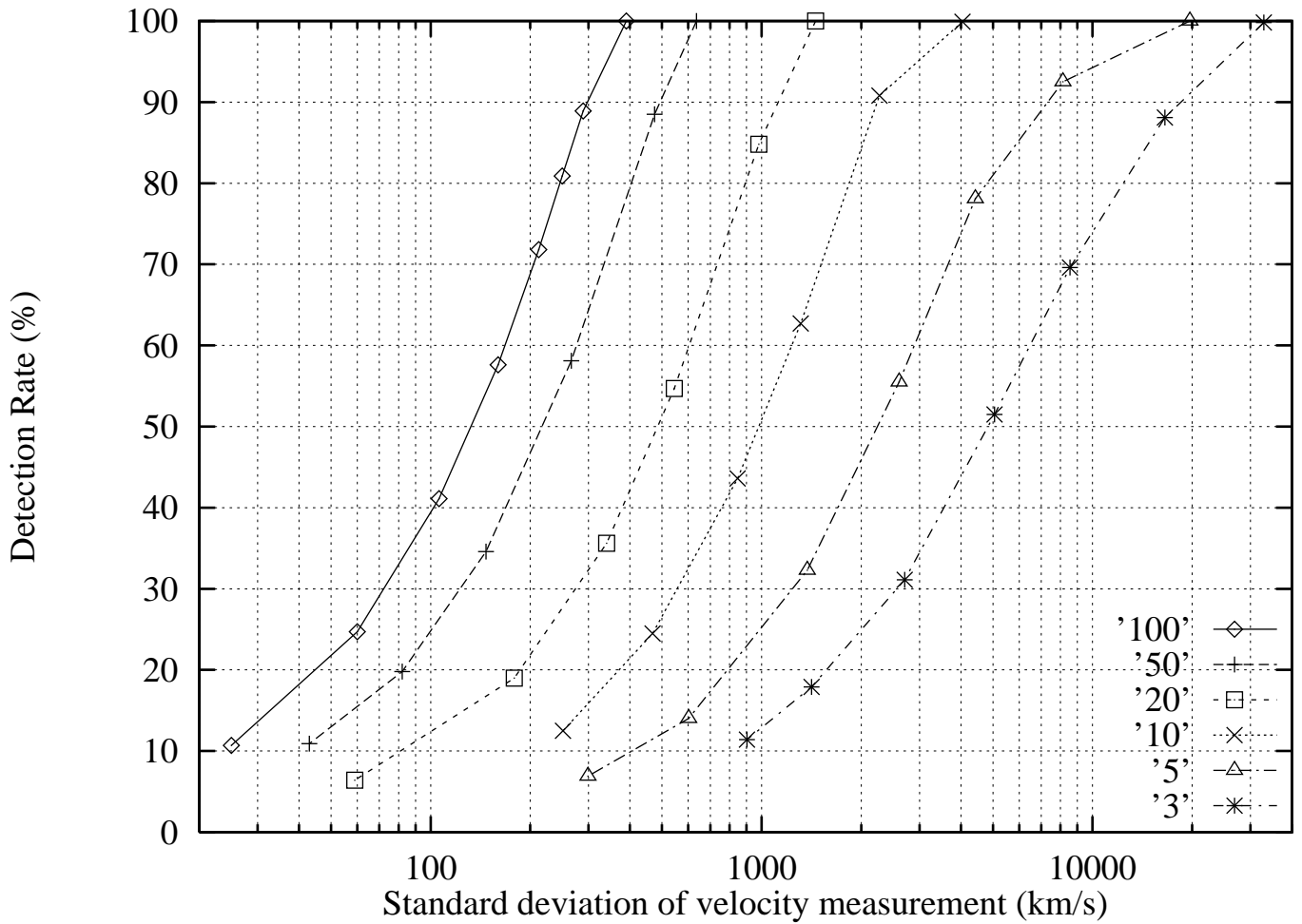

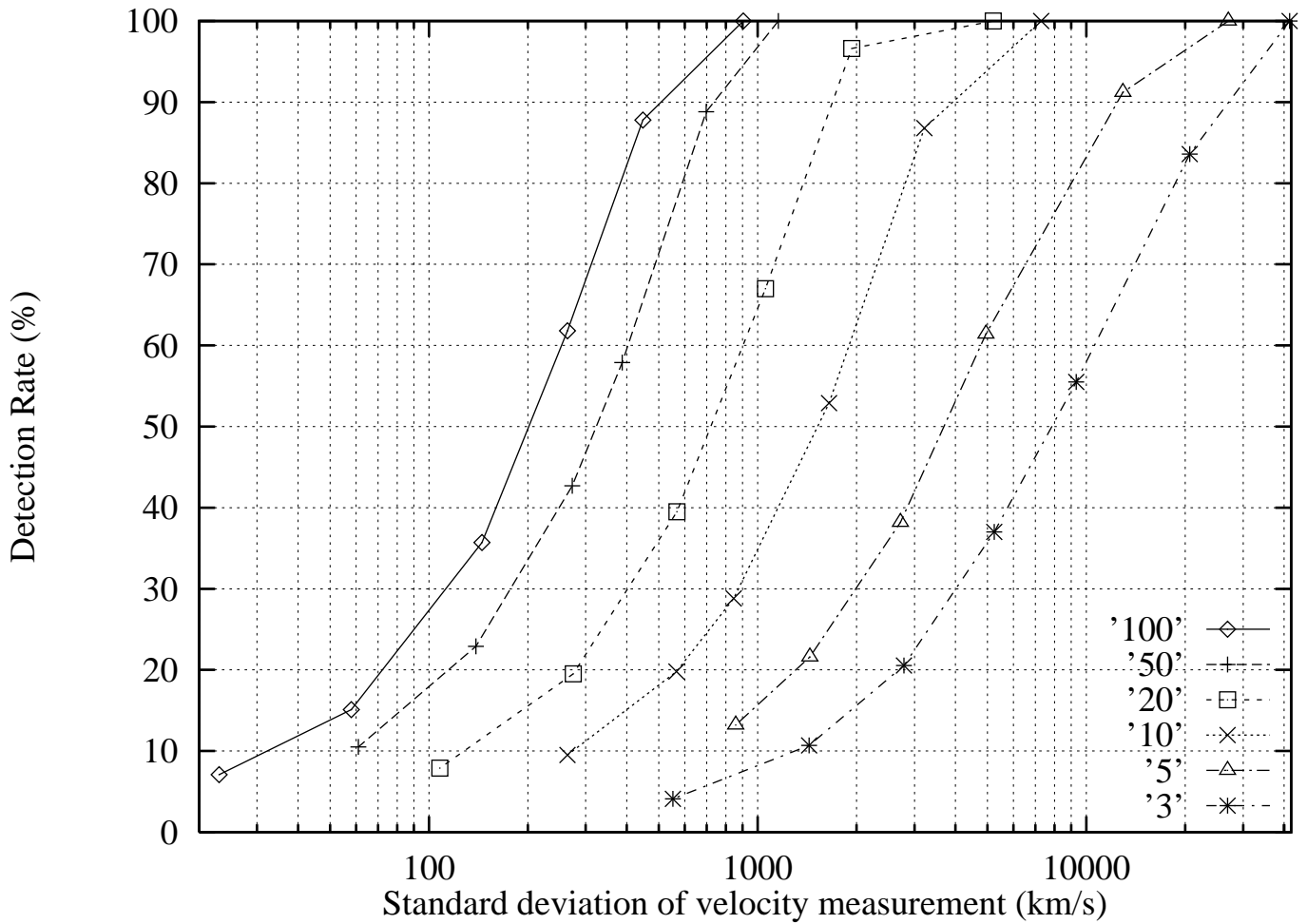

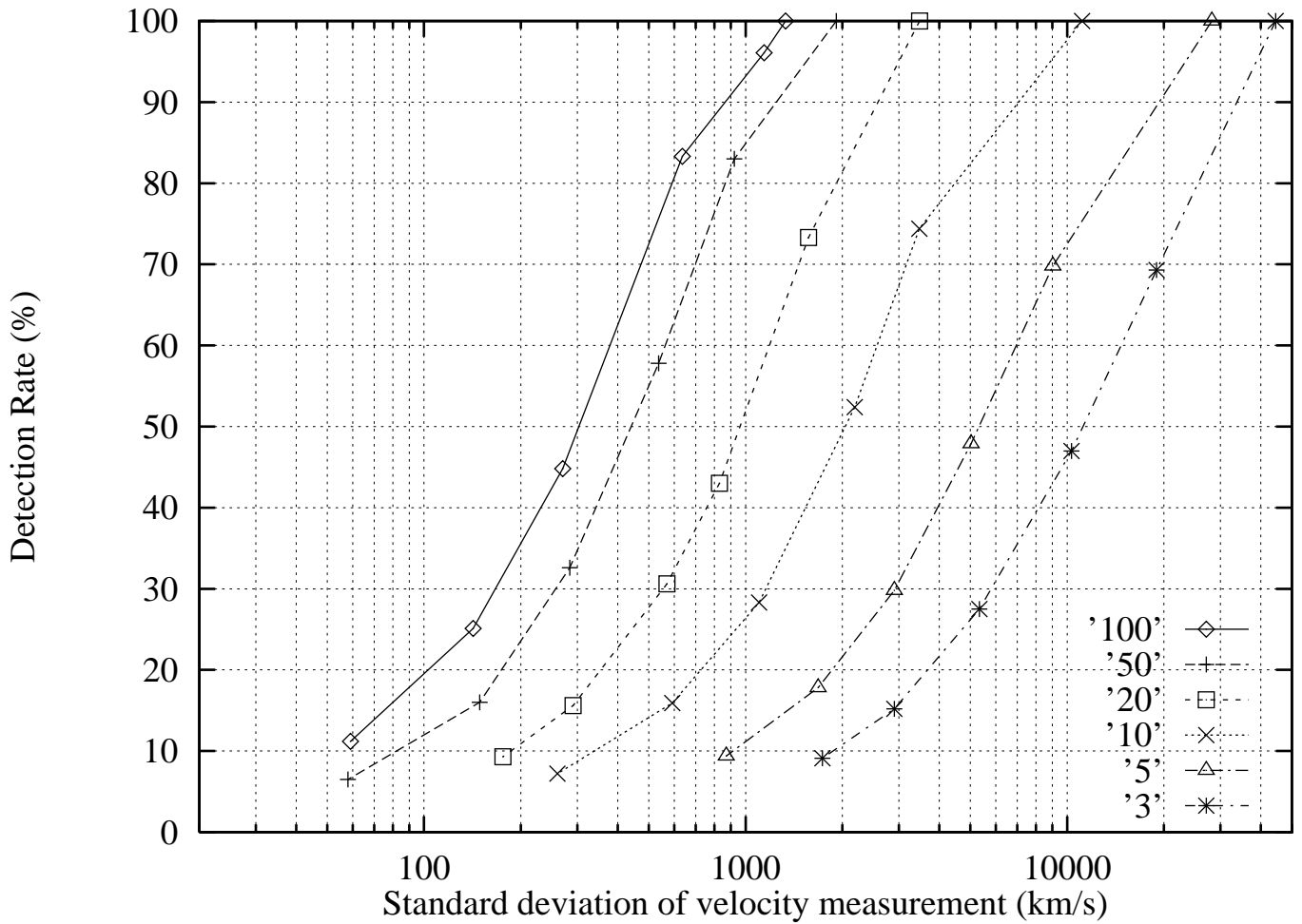

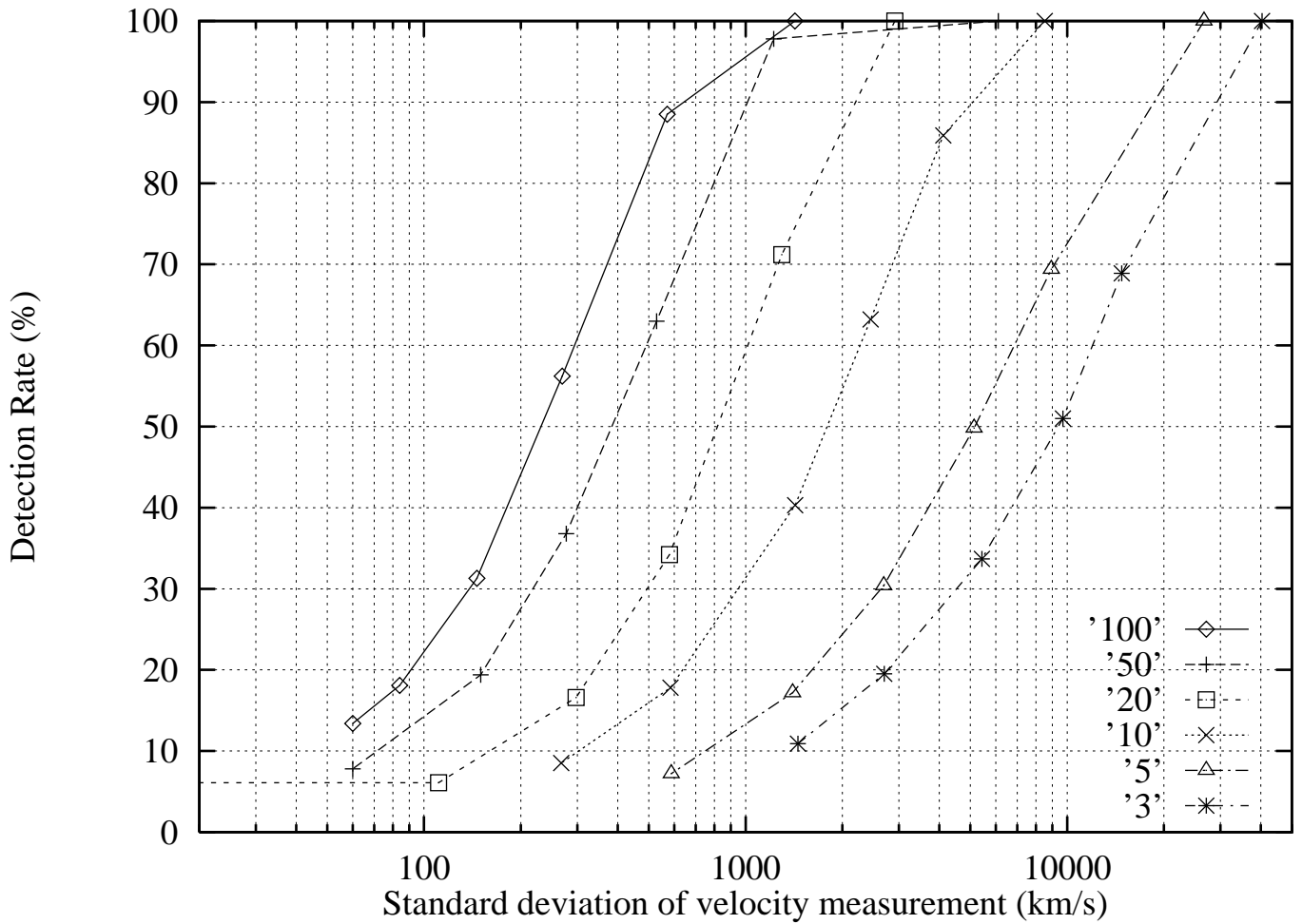

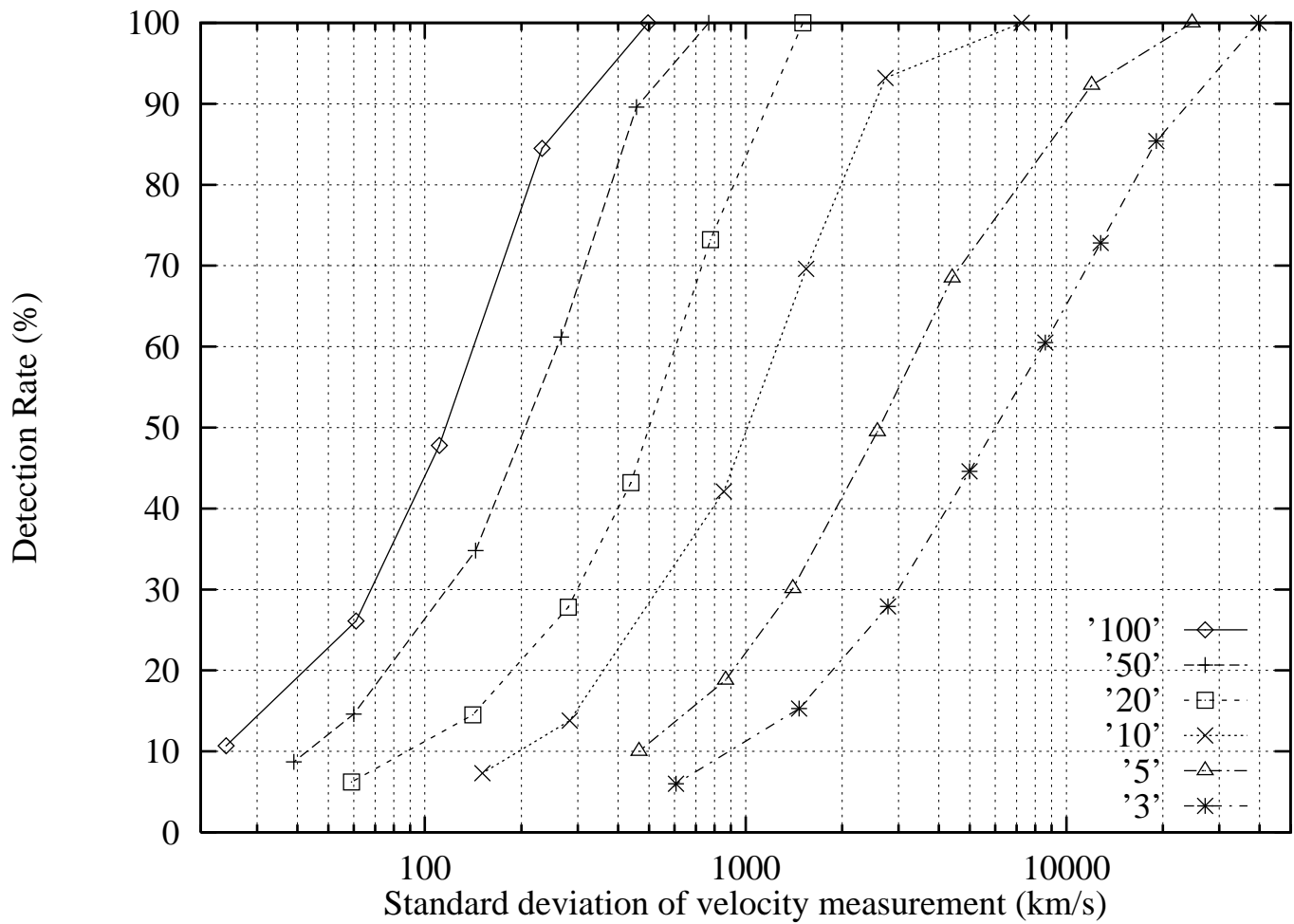

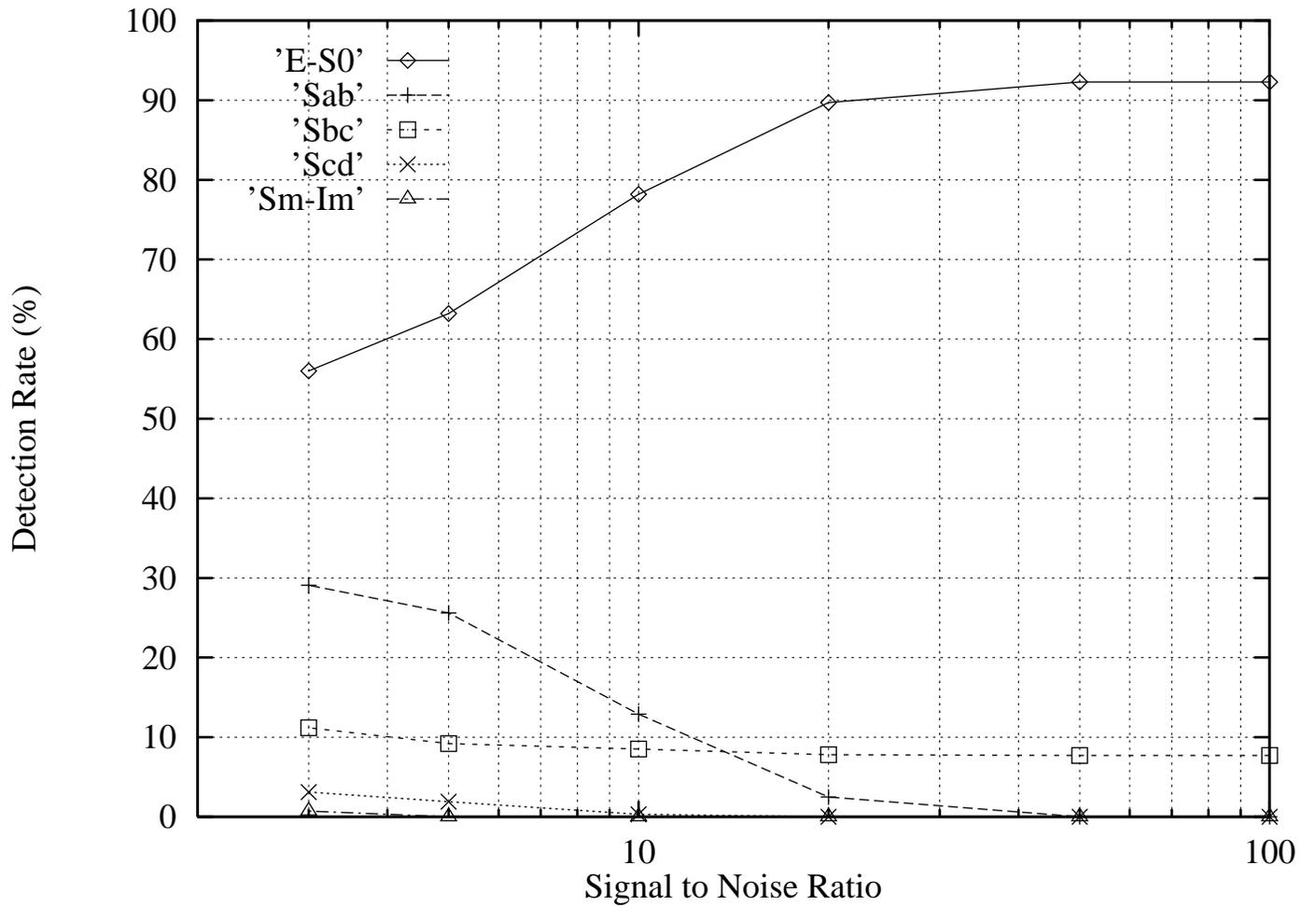

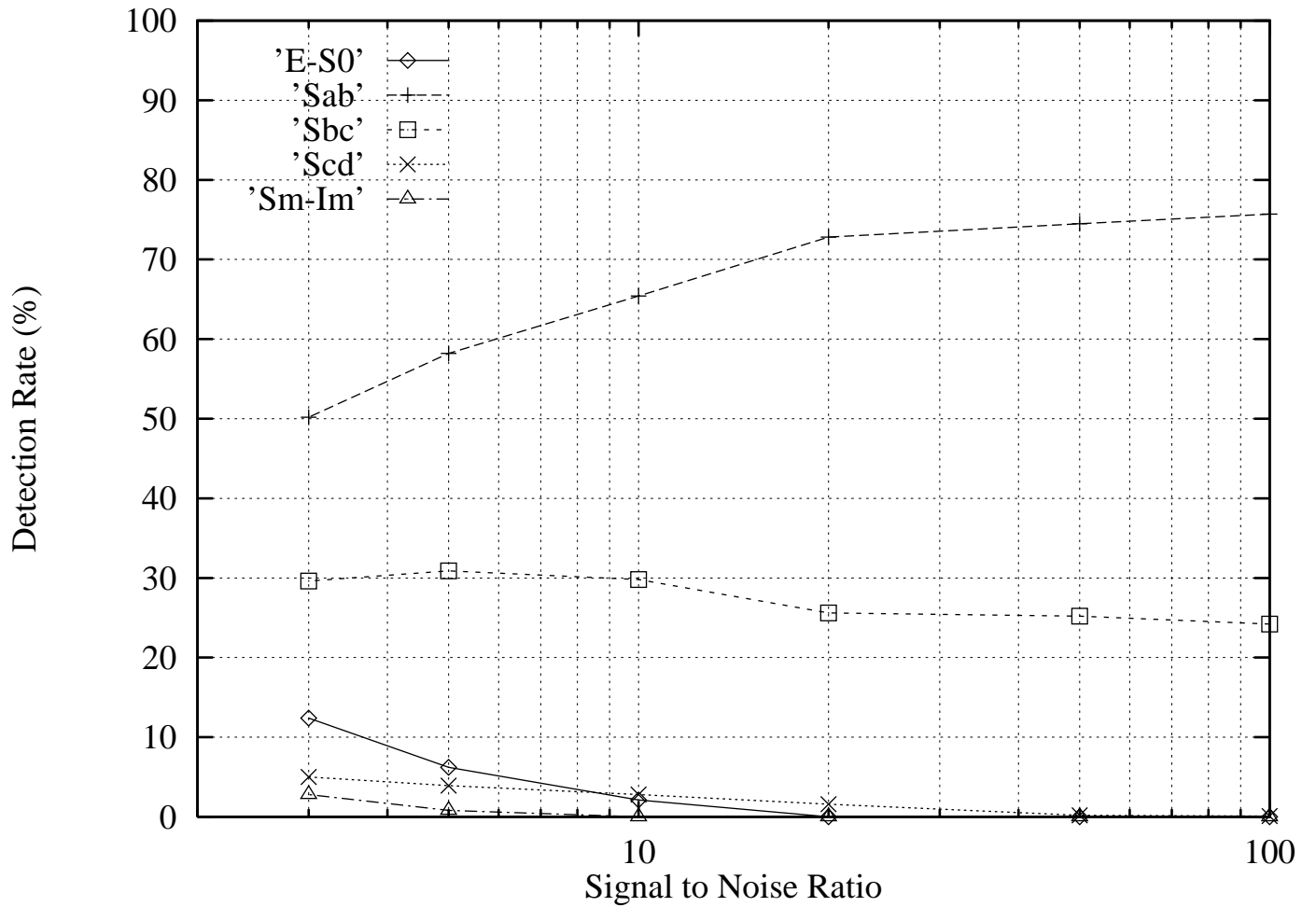

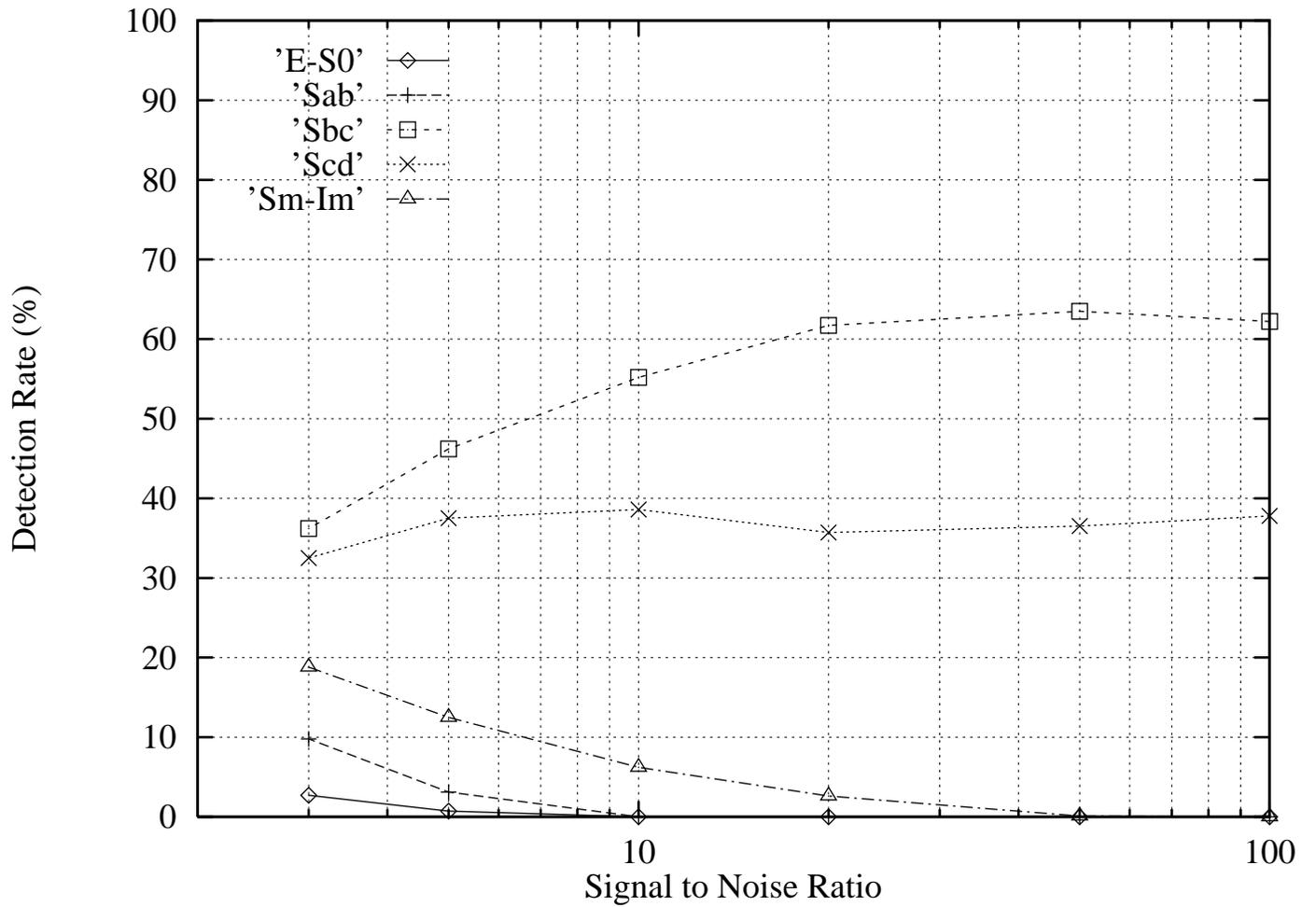

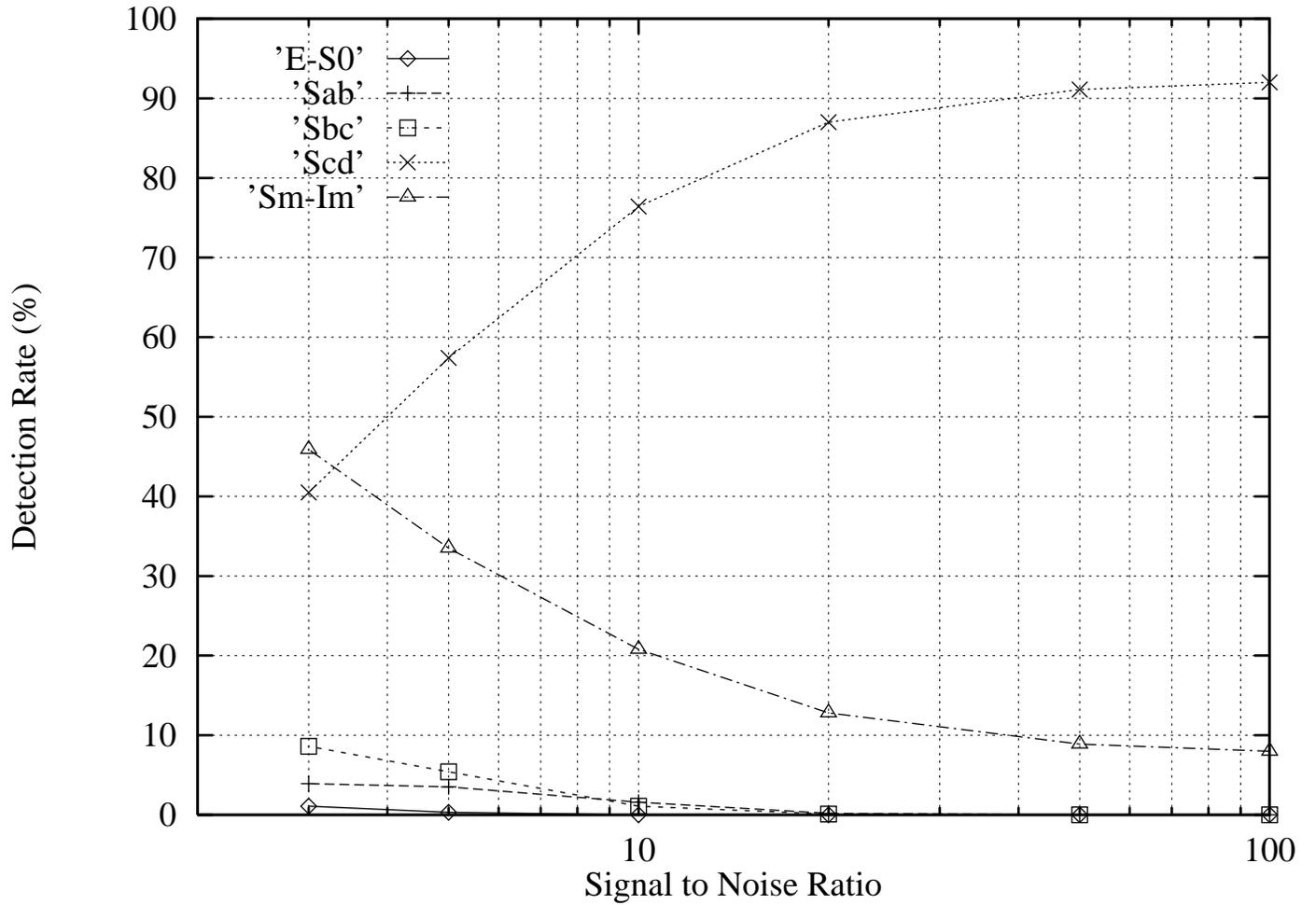

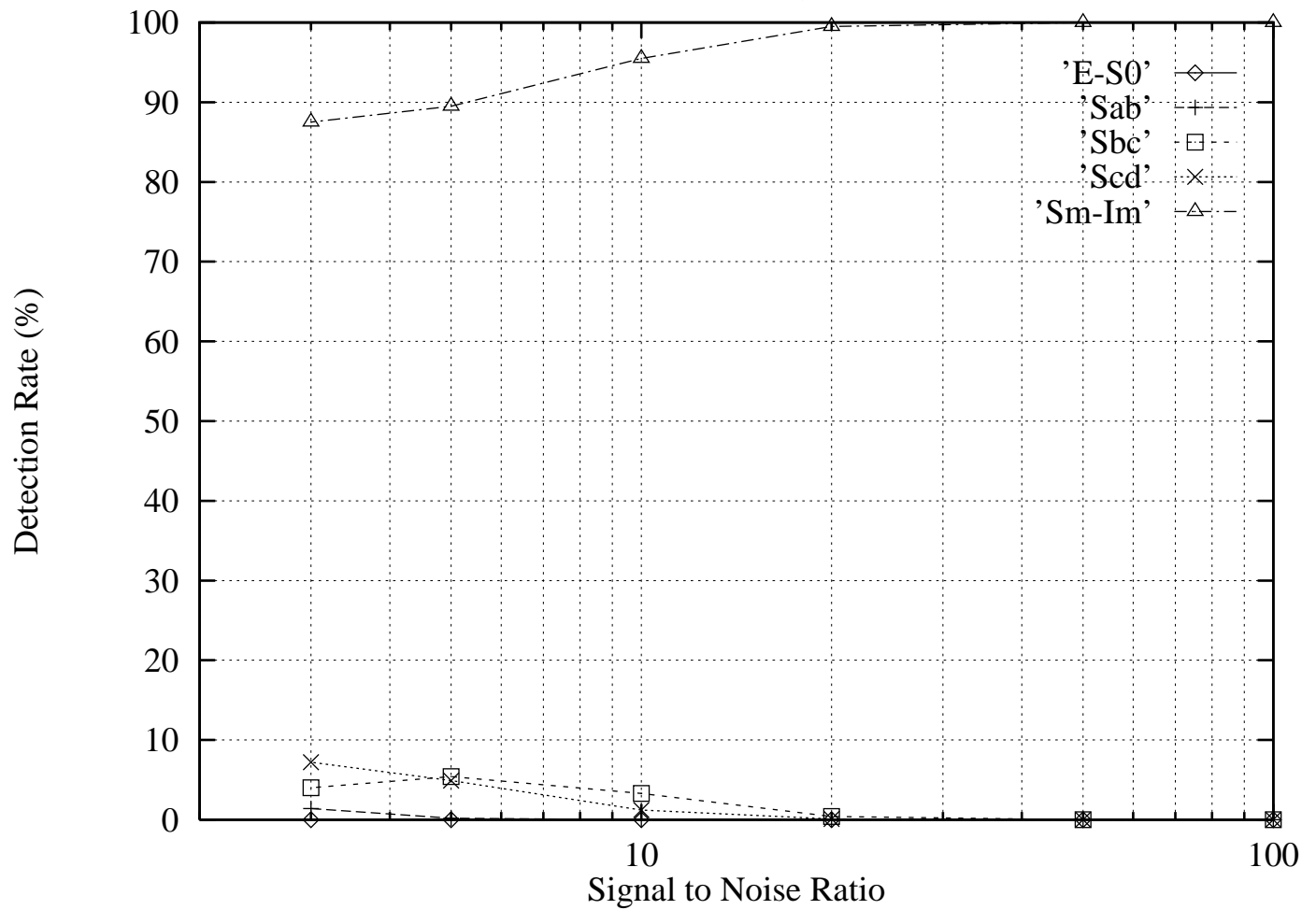

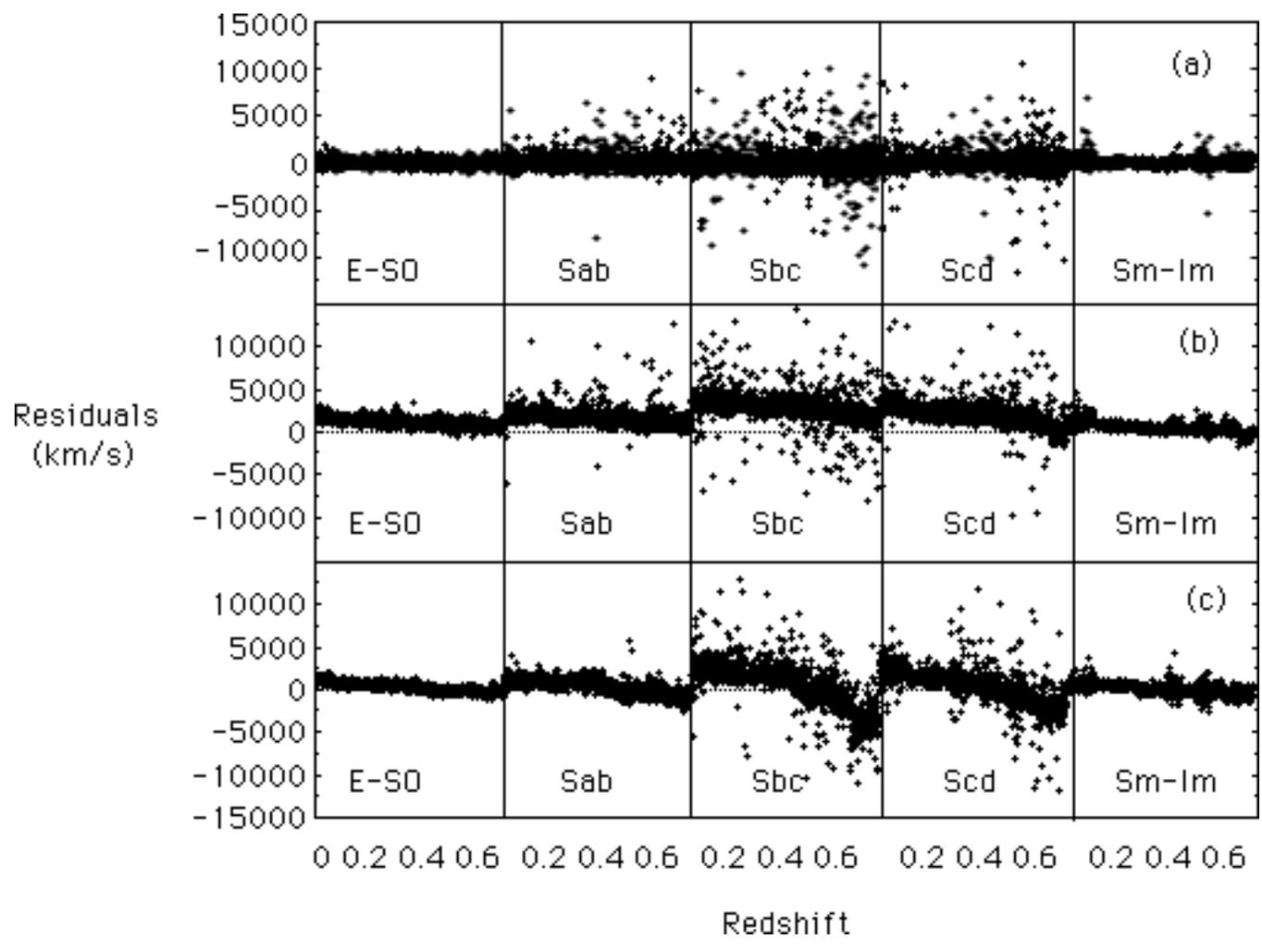

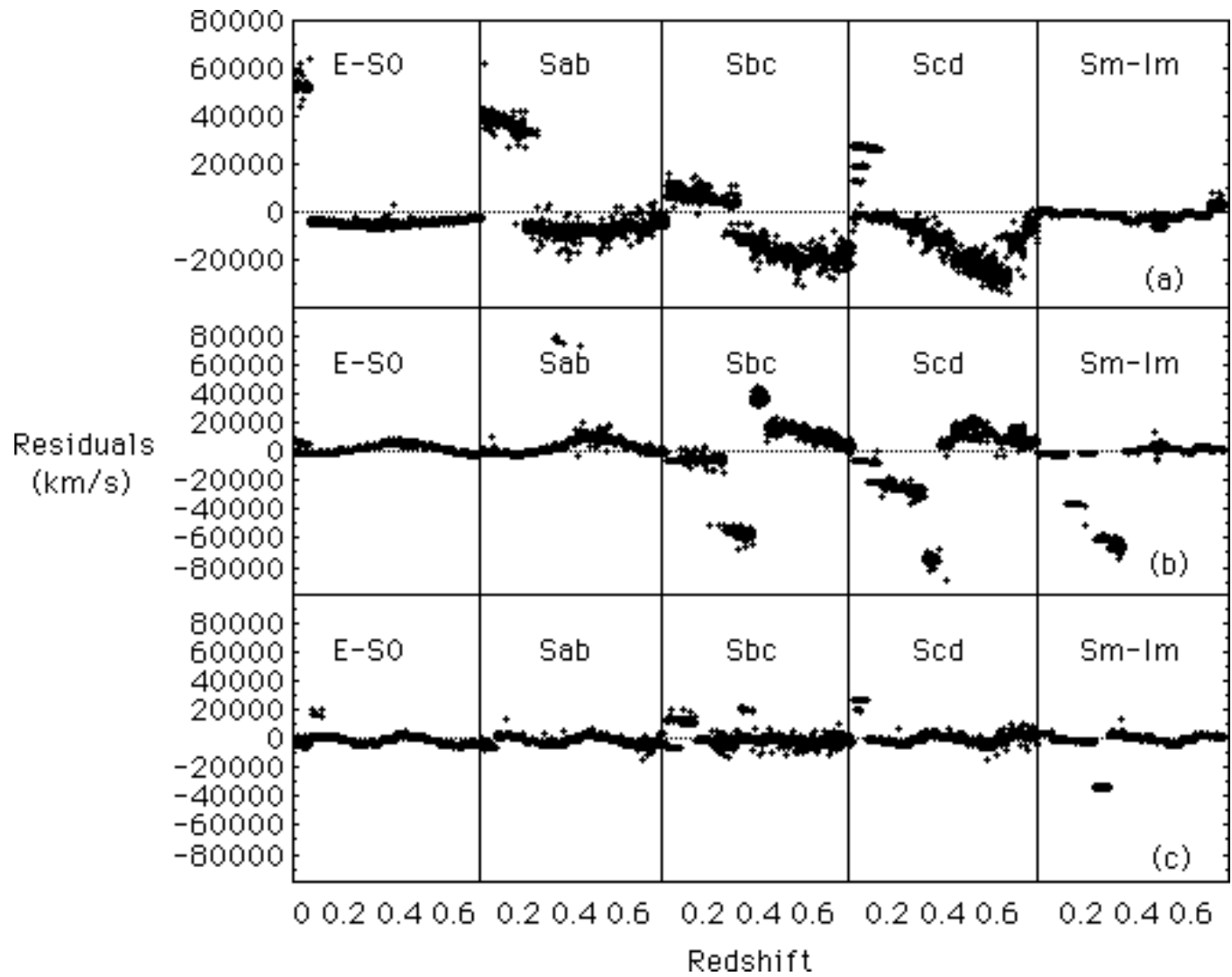